\documentclass[pra,twocolumn,showpacs,letterpaper,showpacs,superscriptaddress]{revtex4-1}
\usepackage{graphicx,amsmath,amssymb,amsfonts,latexsym,color,dcolumn,bm,subfigure}

\begin{document}

\title{Modeling electrostatic patch effects in Casimir force measurements}

\author{R. O. Behunin}
\author{F. Intravaia}
\author{D. A. R. Dalvit}
\affiliation{Theoretical Division, MS B213, Los Alamos National
Laboratory, Los Alamos, New Mexico 87545, USA}
\author{P. A. Maia Neto}
\affiliation{Instituto de F\'{\i}sica, UFRJ, CP 68528, Rio de
Janeiro, RJ, 21941-972, Brazil}
\author{S. Reynaud}
\affiliation{Laboratoire Kastler Brossel, CNRS, ENS, UPMC, Campus
Jussieu, F-75252 Paris France}

\date{ \today}

\begin{abstract}
Electrostatic patch potentials give rise to forces between neutral
conductors at distances in the micrometer range and must be
accounted for in the analysis of Casimir force experiments. In this
paper we develop a quasi-local model for describing random
potentials on metallic surfaces. In contrast to some previously
published results, we find that patches may provide a significant
contribution to the measured signal, and thus may be a more important
systematic effect than was previously anticipated. 
Additionally, patches may render the experimental data at distances below 
1 micrometer compatible with theoretical predictions based on the Drude model.
\end{abstract}


\maketitle

\section{Introduction}

The Casimir effect \cite{Casimir48,Lamoreaux05,Bordag09,Dalvit11} is a remarkable consequence of vacuum field fluctuations
which, in its simplest manifestation, leads to the attraction of two neutral ideal 
conducting plates. At very short distances quantum fluctuation forces dominate the interaction 
between neutral objects making them an essential consideration for micro-electro mechanical
devices (MEMS) and atom traps, among others.
The comparison between experimental measurements and theory for Casimir
forces between metallic plates has been a matter of debate in recent
years. This debate is of particular importance if this comparison is
used to derive constraints on hypothetical new short-range
interactions appearing in addition to the gravity force in
unification models
\cite{Fischbach98,Adelberger03,Onofrio06,Antoniadis11}.

Two recent experiments are at the heart of this debate.
Casimir force measurements by the IUPUI group
\cite{Decca05,Decca07}, performed at distances smaller than 750 nm,
were interpreted by the authors as excluding the dissipative Drude
model and agreeing with the lossless plasma model. This has led to a
discrepancy between experiments and physically motivated theoretical
models, such as the Drude model, for real conductors which exhibit
dissipation. In distinction, a recent experiment by the Yale
\cite{Sushkov11} was able to measure Casimir forces at distances up
to 7 $\mu$m and was interpreted by the authors as being in agreement
with the Drude prediction, including quantum as well as thermal
fluctuations, once an electrostatic patch contribution has been
taken into account.

It is known that patch effects are a source of concern for
Casimir experiments \cite{Speake03,Chumak04,Kim10,deMan10,Kim10b}, as well
as for other precision measurements
\cite{Fairbank67,Camp91,Turchette00,Deslauriers06,Robertson06,Epstein07,%
Pollack08,Adelberger09,Everitt11,Reasenberg11}. For the Yale
experiment the patches were assumed to be much larger than the gap
$D$ between the spherical and planar plates used in the measurement. 
Under
these conditions the patch force is found to be
proportional to $R V^2_\mathrm{rms} / D$ in the proximity force approximation (see below),
where $R$ is the radius of curvature of the spherical plate and $V_\mathrm{rms}$ is
the root-mean-square (rms) voltage of electrostatic patch potentials
\cite{Sushkov11}. For the IUPUI experiment, a patch analysis was
performed with different assumptions leading to the conclusion that
the patch effect had a negligible influence \cite{Decca05}.
Unfortunately, it was not possible in any of these experiments to
measure the patches independently.  It follows that the conclusions
of the theory-experiment comparisons heavily rely on the patch
models used in the data analysis.

In this paper, we revisit electrostatic patch effects and
analyze their possible influence in Casimir force measurements. Our
approach is based on the method pioneered by Speake and Trenkel
\cite{Speake03} with the electrostatic patches described in terms of
a power spectral density. However, we will develop a model for
the power spectral density differing from the one proposed in
\cite{Speake03} and used in \cite{Decca05,Decca07}.

Our model is based on the observation that bare metallic surfaces
are composed of crystallites, each of which constitutes a single
patch, where the local surface voltage is determined by the local
work function \cite{Gaillard06}.
By assuming that through the surface preparation process the
crystallographic orientation, and hence the corresponding work
function, of each crystallite is determined independently and
randomly we can infer that voltage correlations are restricted to
points lying on the same patch: We refer to  this as {\it
quasi-local correlation}.
Our model with quasi-local correlations can be compared to the case
of quenched charge disorder in dielectrics \cite{Naji10,Sarabadani10}, 
and
also shows close similarities with models proposed recently to
describe patch correlation functions for atomic or ionic traps
\cite{Dubessy09,Carter11}.

We will show that  the voltage correlation function from our
quasi-local model strongly differs from that initially proposed in
\cite{Speake03} and used in \cite{Decca05,Decca07}. 
As a result, in contrast to the claims of \cite{Decca05,Decca07}, 
patches may have a significant contribution to the IUPUI measurements. 
%
In addition we will qualitatively address the issue of surface contamination 
which is expected to lead to larger correlation lengths and reduced
voltage fluctuations \cite{Rossi92}. Given that the degree of contamination
is unknown we perform a fit of the patch model we propose to
the difference between measurements and the Casimir theoretical prediction based on the
Drude model and find that it qualitatively explains the residual signal. 
For the Yale experiment, our results will essentially
reproduce those obtained in \cite{Sushkov11}.


\section{Electrostatic patch effect}

In the present section, we recall a few general results of interest,
assuming that the validity conditions of the proximity force
approximation (PFA) are satisfied, that is, the radius of the sphere
used in the experiment is much greater than the sphere-plane
distance. In this case, the expression for the force gradient
$G_{sp}$ (derivative with distance of the force $F_{sp}$) in the
sphere-plane geometry is written as follows in terms of the pressure
$P_{pp}$ (the force per unit area) calculated between two planes
\begin{eqnarray}
\label{GradientPFA}
G_{sp}(D) \equiv \frac{\partial
F_{sp}(D)}{\partial D} = 2 \pi R P_{pp}(D) .
\end{eqnarray}
This expression is used throughout the paper for both Casimir and
patch effects.

The basic description of the patch effect after
\cite{Speake03} is a statistical ensemble of patch potentials
$V_i({\bf r})$ on the surfaces of two planar plates labeled $i=1,2$.
The potentials are assumed to have zero mean $\langle V_i({\bf r})
\rangle =0$, and to be described by the two-point potential
correlation functions
\begin{eqnarray}
\label{defcorrelations} C_{ij}({\bf r}) =  \langle V_i({\bf r})
V_j({\bf 0}) \rangle = \int \frac{d^2 {\bf k}}{4 \pi^2}  e^{ i {\bf
k} \cdot {\bf r} } C_{ij}[\bf k] .
\end{eqnarray}
In the plane-plane geometry, points on the planes are denoted in
cartesian coordinates as ${\bf r} = (x,y)$ and the point ${\bf 0}$
is an arbitrary origin. The correlation functions $C_{ij}({\bf
r})$ and therefore the power spectra $C_{ij}[\bf k]$ are also
assumed to be isotropic. The relations between these two functions
can be written
\begin{eqnarray}
\label{FourierBessel}
&& C_{ij}(r) =  \frac{1}{2\pi} \int_0^\infty   d k \  k \ J_0(k r) \ C_{ij}[k] , \nonumber \\
&& C_{ij}[k] = 2 \pi \int_0^\infty   dr \  r \ J_0(k r) \ C_{ij}(r) ,
\end{eqnarray}
where we have simply denoted $r\equiv\vert{\bf r}\vert$ and
$k\equiv\vert{\bf k}\vert$ and where $J_n(x)$ is the n-th order
Bessel function \cite{Gradshteyn}.
The patch power spectrum $C_{ij}[k]$ corresponds to the notation
$\tilde{C}_{ij}(k)$ in \cite{Speake03}. As usual, the variances and
covariances are given by the integrals
\begin{eqnarray}
\label{covariances} && C_{ij}(0) =  \langle V_i V_j \rangle =
\frac{1}{2\pi} \int_0^\infty  d k \ k \ C_{ij}[k] .
\end{eqnarray}

The pressure due to electrostatic patches in the plane-plane
geometry can be computed exactly \cite{Speake03} as
\begin{eqnarray}
\label{pressure} P_{pp}^\mathrm{patch}(D) &=& \frac{
\varepsilon_o}{4\pi}  \int_{0}^{ \infty }
\frac{ d k \ k^3}{ \sinh^2 (kD)}  \\
&& \times \left\{ C_{11}[k]+ C_{22}[k]- 2 C_{12}[k] \cosh (kD)
\right\} . \nonumber
\end{eqnarray}
It is worth emphasizing at this point that the integral is reduced
to a very simple expression when patch sizes, with a typical value
denoted $\ell_\mathrm{patch}$, are larger than the distance $D$. In
this case, all wavevectors $k$ contributing to the integral
(\ref{pressure}) satisfy $k D \ll 1$, so that the pressure scales
universally as $1/D^2$, irrespective of the particular details of
the power spectrum (Eq. (\ref{covariances}) is used)
\begin{eqnarray}
\label{largepatches} P_{pp}^\mathrm{patch}(D) &=& \frac{
\varepsilon_o}{2D^2}  \int_{0}^{ \infty }
\frac{ d k \ k}{2\pi}   \left\{ C_{11}[k]+ C_{22}[k]- 2 C_{12}[k]  \right\} \nonumber \\
&=&   \frac{ \varepsilon_o}{2D^2} \left\langle \left(V_i -
V_j\right)^2 \right\rangle\quad,\quad D\ll\ell_\mathrm{patch}.
\end{eqnarray}
The above result is expected from the analogy with a
parallel plate capacitor with prescribed voltages. In contrast, when
the relevant wavevectors no longer satisfy the above inequality,
different models for the patch power spectrum result in different
predictions for the patch contribution to the pressure.

It is also worth mentioning here some conditions for the expression
(\ref{pressure}) of the electrostatic patch pressure between two
plates to be valid. A fundamental assumption in this analysis is
that the ergodic hypothesis is satisfied, which means that the
distribution of patches within the interaction area is a fair
approximation of the ensemble-averaged distribution function defined
by the power spectrum $C_{ij}[k]$. When applied to two plane plates
of finite area $A$, we expect this assumption to be well satisfied
if the effective interaction area contains a large number of patch
correlation areas $A\gg\ell_\mathrm{patch}^2$. For the sphere-plane
geometry, the effective area of interaction is of the order of
$\pi DR$, leading to the validity requirement
\begin{eqnarray}
\label{validity} \pi D R \gg \ell_\mathrm{patch}^2 .
\end{eqnarray}

In the following two subsections we recall a model used in
\cite{Speake03} and \cite{Decca05,Decca07}, and introduce
another model with quasi-local correlations which we think to be a
better description of sputtered surfaces.


\subsection{The sharp-cutoff model}

We now discuss the model of patch correlations which was
proposed as an example in \cite{Speake03} and then used in
\cite{Decca05,Decca07} to assess the contribution of electrostatic patches
to the Casimir force measurements.

It is a simple description based upon two assumptions: a) the power
spectrum of patches is an annulus in $k$-space possessing no other
dependence than a sharp cutoff at small ($k_\mathrm{min}$) and large
($k_\mathrm{max}$) wavevectors (hence the name {\it sharp-cutoff
model}); b) there are no cross correlations between the two plates
($C_{12} = 0$). This model gives the power spectrum for a single
plate as
\begin{equation}
\label{STPS} C_{ii}[k] = \frac{ 4\pi  V_\mathrm{rms}^2}{
k_\mathrm{max}^2 - k_\mathrm{min}^2 } \theta(k_\mathrm{max} - k)
\theta( k - k_\mathrm{min}),
\end{equation}
where $V_\mathrm{rms}^2$ is the variance of the potential on one
plate and $\theta$ is the Heaviside step function.

In order to determine the parameters of this model, the authors of
\cite{Decca05} used the further assumptions~: c) based on AFM images
of the surfaces, the minimum and maximum grain sizes of the samples
were determined to be $\ell_\mathrm{patch}^\mathrm{min}=25$ nm  and
$\ell_\mathrm{patch}^\mathrm{max}=300$ nm ~; d) the patch sizes were
assumed to be the same as the grain sizes and the cutoffs in
$k$-space were derived from the inverse maximum and minimum grain sizes
$k_\mathrm{min}=2\pi/\ell_\mathrm{patch}^\mathrm{max}=20.9 \mu\mathrm{m}^{-1}$ and
$k_\mathrm{max}=2\pi/\ell_\mathrm{patch}^\mathrm{min}=251 \mu\mathrm{m}^{-1}$ ~; e) the rms voltage was obtained by computing
the variance of the work functions over the different
crystallographic planes of gold, which led to $V_\mathrm{rms} \approx
80.8$ mV. Using the five assumptions a) to e), it was concluded in
\cite{Decca05} that the patch pressure had a negligible influence on
the estimation of the Casimir force. A reasonable agreement was then
obtained between the experimental data and the prediction for the
Casimir pressure using the lossless plasma model (more discussions
below).

\begin{figure}[t]
\centering
\includegraphics[width=2.7in]{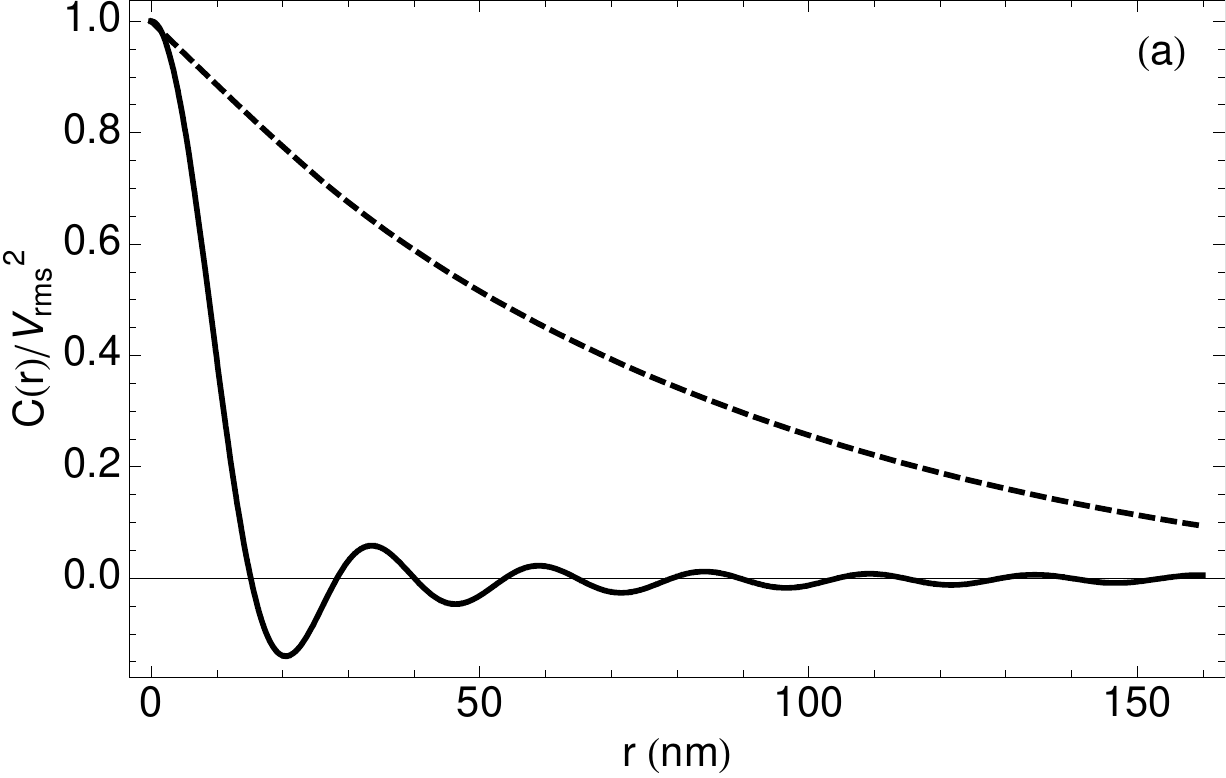}
\includegraphics[width=2.7in]{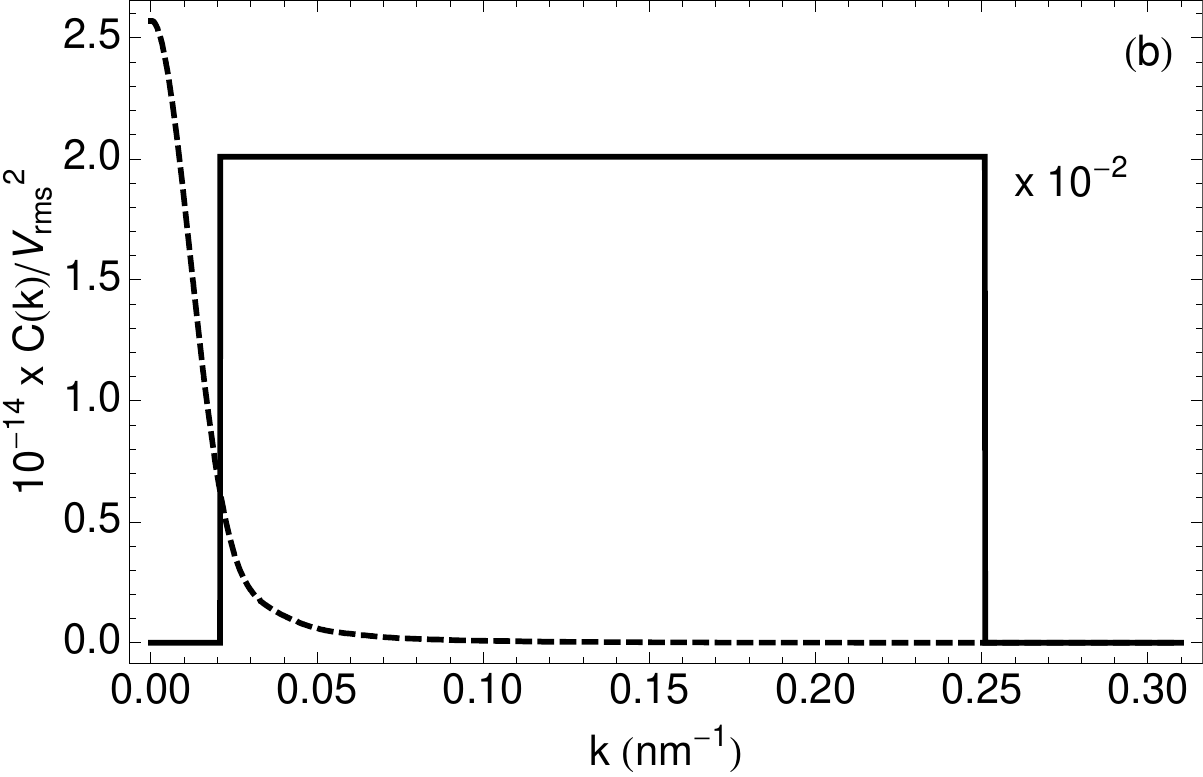}
\caption{Comparison of the sharp-cutoff and quasi-local patch models
described in subsections II-A (dashed lines) and II-B (solid lines),
respectively. Plot (a) shows the voltage correlation functions in
real space while plot (b) shows the associated spectrum in Fourier
space. All plots correspond to the correlation function $C \equiv C_{ii}$ divided
by $V_\mathrm{rms}^2$. On the lower plot, the sharp-cutoff spectrum
discussed in II-A is multiplied by a factor of $100$ in order for it
to appear at the scales shown. The parameters used for both models,
discussed in subsections II-A and II-B, are taken from
\cite{Decca05}, but however, do not correspond to the same average
patch size.} \label{correlation_functions}
\end{figure}

Now we will argue that model (\ref{STPS}) is not a good
description for  the patch power spectrum for
the surfaces used in the experiments, and later on, we
will also question the relation between patch and grain sizes. In
order to make the former point clear, let us write the correlation
function $C_{ii}(r)$ of patches in real space which can be obtained
through an inverse Fourier transform (\ref{FourierBessel}) from the
spectrum (\ref{STPS})
\begin{eqnarray}
\label{ST_Spatial_Correlation} C_{ii}(r) = 2 V_\mathrm{rms}^2
\frac{k_\mathrm{max} J_1( k_\mathrm{max} r )-k_\mathrm{min} J_1(
k_\mathrm{min} r )}{\left(k_\mathrm{max}^2   -
k_\mathrm{min}^2\right)r} .&&
\end{eqnarray}
As one moves away from coincidence the correlation function
$C_{ii}(r)$ oscillates between positive and negative values with a
period of the order of the smallest patch size, and is contained
within an envelope decaying as $r^{-3/2}$ (see
Fig.\ref{correlation_functions}). These oscillations imply that the
patch potential shows {\it correlations} as well as {\it
anti-correlations} in space. Such behavior could be expected for
surfaces exhibiting some kind of antiferroelectric ordering (where
the configurational energy is minimized when adjacent surface
dipoles are antiparallel), but will unlikely describe the random
potentials on sputtered surfaces.

As already stated, the strict relation between patch sizes and grain
sizes, assumed in the analysis of \cite{Decca05}, has also to be
questioned. The adsorption of contaminants on the surfaces
alters patch sizes which, as a result, do not necessarily correspond
to the grain sizes \cite{Rossi92}. We expect that contamination
leads to an effective smearing of the patch layout, so that patch
sizes will be larger than grain sizes while the voltage variance
will be less than the value obtained for a clean sample from the
assumption e) discussed above. 



\subsection{The quasi-local correlation model}

We now propose another patch model which we think to be a better
motivated description of the patch correlation function for the
surfaces used in the experiments.

To model the layout of cystallites on a plate, we choose a random
patch layout and afterward assign a random potential to each patch.
For a given micro-realization of patches we  write the voltage over
the whole surface as
\begin{equation}
\label{voltage}
V( {\bf x} ) = \sum_a v_a \Theta_a ({\bf x}).
\end{equation}
The sum is over all patches, $v_a$ is a random variable describing
the voltage on patch $a$, and the function $\Theta_a( {\bf x})$ is
defined to be 1 for ${\bf x}$ on the $a$th patch, and 0 otherwise.

We now obtain the two-point voltage correlation function by
performing ensemble averages over all micro-realizations of the
patch voltages and layouts. Physically, the voltage on each site is
determined by the crystallite face exposed to the surface. As we
assume that each crystallite is deposited with a random
crystallographic orientation and that each deposition is
statistically independent we can infer that
\begin{equation}
\label{variance} \langle v_a v_b \rangle_v = \delta_{ab}
V^2_\mathrm{rms},
\end{equation}
where the expectation value $\langle ... \rangle_v$ averages over
the voltage fluctuations only and $\delta_{ab}$ is the Kronecker delta. 
Also, note that we are implicitly assuming that there are no cross correlations
between the patches on different plates, $C_{12} = 0$.
Using (\ref{voltage}) and
(\ref{variance}) we construct the two-point voltage correlation for
{\it a single micro-realization} of the patch layout
\begin{equation}
\label{QLPatchModel}
 \left\langle  V({\bf x} ) V({\bf x'}) \right\rangle_v =
 V^2_\mathrm{rms} \sum_a \Theta_a({\bf x} ) \Theta_a({ \bf x'}).
\end{equation}

\begin{figure}[h]
\begin{center}
\includegraphics[width=2in]{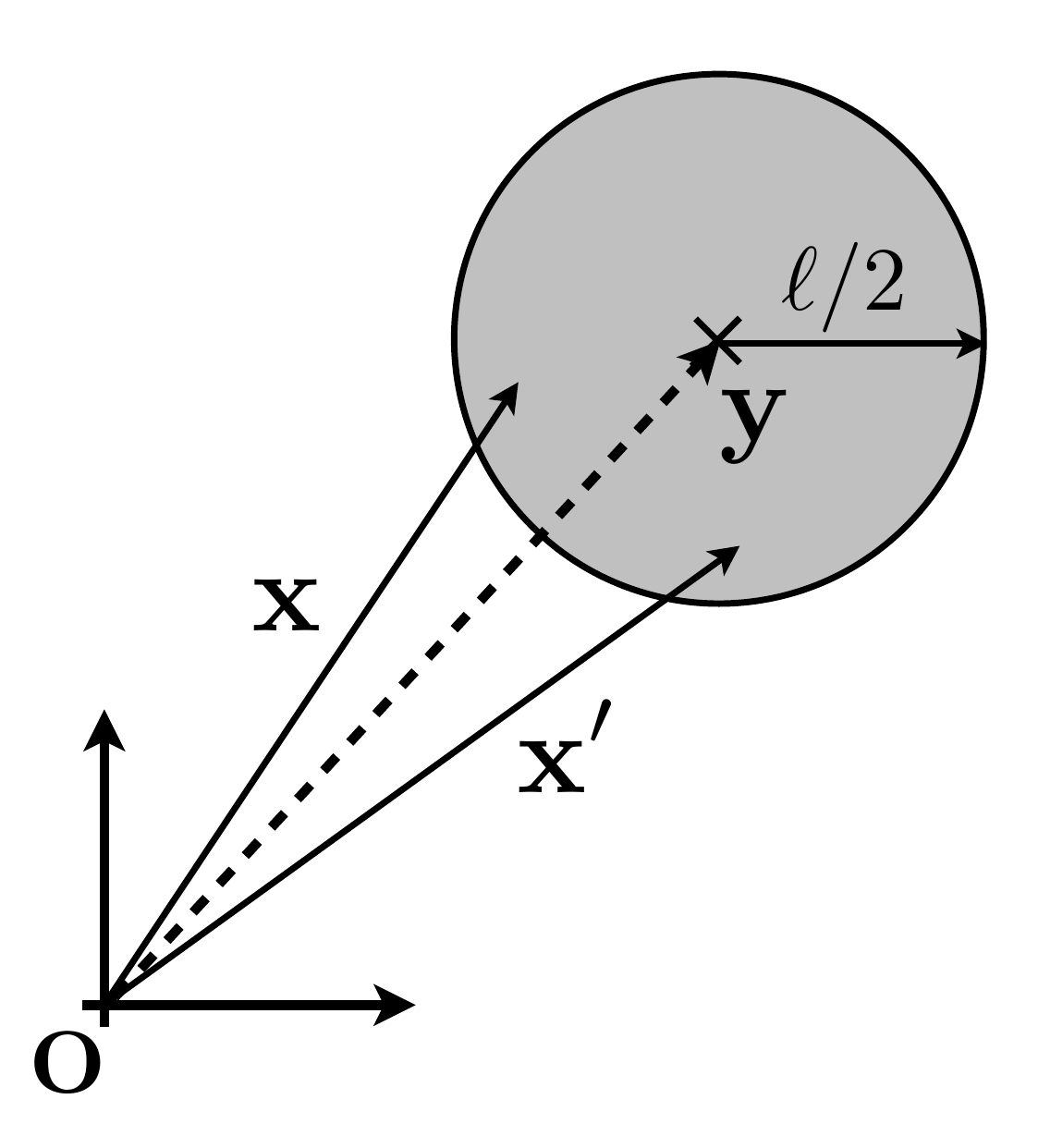}
\caption{The voltage correlation function, $C({\bf x}, {\bf x}')$,
is constructed by summing over all circular patches which contain
both ${\bf x}$ and ${\bf x}'$. This is undertaken by integrating
over patch centers ${\bf y}$, and accounting for the distribution in
patch sizes with the distribution $\Pi(\ell)$. }
\label{PatchCorrelation}
\end{center}
\end{figure}

The final step in constructing the ensemble-averaged voltage
correlation function  is to average over all patch layouts. We carry
this out by exploiting several symmetries:

 \begin{enumerate}
 \item We assume that the patches are distributed uniformly and
 isotropically which implies that the average patch associated
 with any given point on the surface is circular with a radius
 determined from a distribution of patch sizes. In reality no
 patch is circular and this notion of {\it patch radius} should
 only be taken in a statistical sense.

 \item For any two points on the sample surface, the voltage
 correlation function $C({\bf x}, {\bf x}')$ is proportional to
 the number of patches which contain both points (among all micro-realizations).
 By employing  the statistical description of patches, as described above in 1,
 the correlation will be computed by  summing over all circular
 patch centers and sizes as depicted in Fig. \ref{PatchCorrelation}.

 \item As a check one can verify that the correlation at
 coincidence is the constant $V_\mathrm{rms}^2$. Moreover, translational
 and rotational invariance implies that $C({\bf x}, {\bf x}')$
 depends only on $r = |{\bf x} - {\bf x}'|$.

 \end{enumerate}

Given these considerations we find the following form for the
correlation function:
\begin{eqnarray}
C({\bf x}, {\bf x}')  &&=   \int_0^\infty d\ell \ \Pi(\ell) \\
&& \times \frac{4 V^2_\mathrm{rms} }{\pi \ell^2 } \int  d^2 y \
\theta( \ell/2 - | {\bf x} - {\bf y}|)  \theta( \ell/2 - | {\bf x}'
- {\bf y}|), \nonumber
\end{eqnarray}
where the integral over $y$, constrained by the $\theta$-functions,
sums over all patches of size $\ell$ which contain both points. The
final integral over $\ell$ averages over patch sizes where
$\Pi(\ell)$ is the distribution of patch diameters. Note in
particular that the translational invariance of the correlation
function is made apparent by the change of variables ${\bf x}' -
{\bf y} \to {\bf z}$ . Subsequently performing the integration over
$y$ reveals the rotational invariance of the final result
\begin{eqnarray}
\label{quasi-local-correlation}  C(r)  \equiv C_{ii}(r) &&=   \frac{ 2
V^2_\mathrm{rms}}{\pi}  \int_r^\infty d\ell \ \Pi(\ell)
\nonumber \\
&& \times \bigg[ \cos^{-1}\left(\frac{r}{\ell} \right) -
\frac{r}{\ell}\sqrt{ 1 -\left( \frac{r}{\ell} \right)^2 } \bigg].
\end{eqnarray}

The patch power spectrum can then be obtained through the Fourier
transform (\ref{FourierBessel}). Some interesting properties can be
given at this point. First, the integration of the correlation
function over all space is simply
\begin{eqnarray}
\label{QLCorrelationK0} C[k=0] = \frac{1}{4} \pi \overline{\ell^2}
V^2_\mathrm{rms},
\end{eqnarray}
with $\overline{\ell^2}$ the variance of the distribution
$\Pi(\ell)$. Second, the integral of $C[k]$ over wavevectors is just
the variance of the potential $\int_0^\infty dk k C[k] = 2\pi
V^2_\mathrm{rms}$.

One can derive some universal scaling laws for the patch
contribution  to the pressure in some limiting cases. When the
patches are much larger than the gap ($D\ll \bar{\ell}$), the
expression (\ref{largepatches}) is obtained. This $1/D^2$ scaling
law for the pressure (and the corresponding $1/D$ for the energy per
unit area for planar plates) is universal for all patch power
spectral densities whenever the typical patch sizes are much larger
than the gap. In particular, this scaling was used in
\cite{Sushkov11} to model the patch effect \cite{footnote}. In the
opposite limit, where the typical patch sizes are much smaller than
the gap, one can obtain a simple scaling law. In this case the
spectrum $C[k]$ is approximately constant over the wavevector range
$k\stackrel{<}{\scriptscriptstyle{\sim}}1/D$ which provides the most
significant contribution to (\ref{pressure}).  We then find, when
using (\ref{QLCorrelationK0}),
\begin{eqnarray}
\label{long-distance} P^\mathrm{patch}(D) & \simeq &
\frac{\varepsilon_0}{2\pi}\, C[0]\,
\int_0^{\infty}  dk\,\frac{k^3}{{\rm sinh}^2(kD)} \\
&\simeq&  \frac{3\zeta(3)}{4} \frac{\varepsilon_0 V_\mathrm{rms}^2
\overline{\ell^2}}{D^4} \approx 0.90 \frac{\varepsilon_0
V_\mathrm{rms}^2 \overline{\ell^2}}{D^4}. \nonumber
\end{eqnarray}
We emphasize at this point that this $1/D^4$ scaling law is generic
for all spectra having a finite limit at $k=0$, but does not hold
when $C[k]$ vanishes at $k=0$. In particular, in the model discussed
in II-A, there is a sharp-cutoff of the power spectral density at
$k_\mathrm{min}>0$. In this case, the pressure (\ref{pressure}) is
exponentially small when $k_\mathrm{min} D\gg 1$, that is also $D\gg
\ell_\mathrm{patch}^\mathrm{max}$. The leading order contribution
indeed comes from the exponential tail of $1/\sinh^2(kD)$ and is
much smaller than the result found in the generic case
(\ref{long-distance}). This point will play a crucial role in the
comparison to experimental data discussed in the next section.

Before entering this discussion we  choose a specific form for the
patch size distribution $\Pi(\ell)$ which is similar in spirit to
the sharp-cutoff model discussed in subsection II-A. By assuming the patch
sizes are distributed uniformly within a finite interval between a
minimum $\ell_\mathrm{patch}^\mathrm{min}$ and maximum
$\ell_\mathrm{patch}^\mathrm{max}$ value, the probability
distribution is
\begin{equation}
\label{SM} \Pi(\ell) = \frac{\theta(\ell_\mathrm{patch}^\mathrm{max}
- \ell) \theta(\ell - \ell_\mathrm{patch}^\mathrm{min} ) }{
\ell_\mathrm{patch}^\mathrm{max}- \ell_\mathrm{patch}^\mathrm{min} }
,
\end{equation}
and has the following moments
\begin{eqnarray}
\label{SMmoments} &&\overline{\ell} =
\frac{\ell_\mathrm{patch}^\mathrm{max} +
\ell_\mathrm{patch}^\mathrm{min} }{2} \\
&&\overline{\ell^2} = \frac{(\ell_\mathrm{patch}^\mathrm{max})^2 +
(\ell_\mathrm{patch}^\mathrm{min})^2
+\ell_\mathrm{patch}^\mathrm{max} \ell_\mathrm{patch}^\mathrm{min}
}{3}. \nonumber
\end{eqnarray}
Additionally, we would like to remark that
we have also considered other size patch size distributions $\Pi(\ell)$ (log-normal,
Gaussian, generalized gamma, etc.) and have found similar results
for the pressure in all cases.

We emphasize that, despite some similarity in the construction of
the two models discussed in subsections II-A and II-B, they
correspond to very different correlation properties, the most
striking difference resulting from a nonvanishing value for $C[k=0]$
in the quasi-local model which gives a distinct large distance behavior. 
In particular, a patch model employing quasi-local correlations was 
recently adopted to describe heating in ion traps and dissipation 
in cantilevers \cite{Dubessy09}. There, the observed large distance 
$(D \gg \overline{\ell})$ scaling of electric field noise 
($\propto D^{-4}$) is linked with a nonvanishing value of $C[0]$.  

To estimate the effects of contamination, we will assume
that the patch power spectrum on a dirty surface
takes the same form as on a clean surface (i.e., also given by the quasi-local model), with the exception
that the parameters of the model are altered by
the contaminants. 
Let us stress that quasi-local correlations may
be not as accurate for contaminated surfaces as for clean ones.
We employ the above assumptions in a preliminary manner to account for the properties
of contaminated surfaces, to be confirmed by dedicated studies
to come in the future.

\section{Comparison with experiments}

We now compare the theory and experiments by calculating the Casimir
force from the Drude model, and the patch pressure arising from the
model with quasi-local correlations. To make the comparison we first
calculate the plane-plane Casimir pressure $P_{pp}(D)$ at temperature $T$ using the Lifshitz formula 
\cite{Lifshitz,Lambrecht00,Lambrecht06}.
We use tabulated optical data for gold \cite{Palik}, 
extrapolated to low frequencies with a Drude model to describe the contribution of conduction electrons, 
$\varepsilon_{\rm cond}(\omega) = 1 - \Omega_P^2 / (\omega (\omega+i \gamma))$,
where $\Omega_P$ is the plasma frequency and $\gamma$ quantifies the damping rate.
To account for 
roughness corrections to
the Casimir pressure
we adopt the 
simplest formulation based on an additive scheme (Eq. (33)
in \cite{Decca05}).
We will call the resulting pressure as the ``Drude model"  
Casimir pressure $P_{pp}^{\rm Drude}(D)$.

As already stated, we use the PFA to relate the experimental data
corresponding to the sphere-plane geometry to the predictions
calculated in the plane-plane geometry, for the Casimir and the
patch effects. In the IUPUI experiments the sphere-plane force
gradient $G_{sp}$ is measured, which is related to the equivalent
plane-plane pressure as in (\ref{GradientPFA}). In the Yale
experiments the sphere-plane force $F_{sp}$ is measured, which is
related similarly to the plane-plane energy per unit area.

After subtracting from the experimental data the theoretical
predictions for the Casimir interaction, we find a residual signal
\begin{equation}
\label{residuals} 
\delta{P}^{\rm Drude} (D) \equiv P^{\rm
experiment}_{pp}(D) - P^{\rm Drude}_{pp}(D) .
\end{equation}
The question we address in the following is whether or not the
residual $\delta{P}^{\rm Drude}$ can be explained by a reasonable modeling of
patch effects. The criterium is then to minimize the remaining
difference between the residual signal and the patch pressure
$\delta{P}^{\rm Drude} (D) - P^\mathrm{patch} (D)$. The residual is
defined here for the Drude model and may be as well be defined for
the plasma model. The patch pressure $P^\mathrm{patch} (D)$ is then
defined for a given patch model, say in particular the sharp-cutoff
(subsection II-A) or quasi-local (subsection II-B) models.


\subsection{Data analysis for the IUPUI experiment}

For the comparison with the IUPUI experiment we compute the Casimir
force at room temperature $T=295$ K using tabulated optical data extrapolated to low frequencies with a Drude model 
with parameters $\Omega_P = 8.9$ eV for the plasma frequency and $\gamma
= 0.0357$ eV for the damping rate.
 Root mean square roughness heights for the plane and the sphere
 are $3.6$ nm and $1.9$ nm, respectively.
These permittivity and roughness parameters are the ones reported in \cite{Decca07}. 

\begin{figure}[t]
\centering
\includegraphics[width=3.3in]{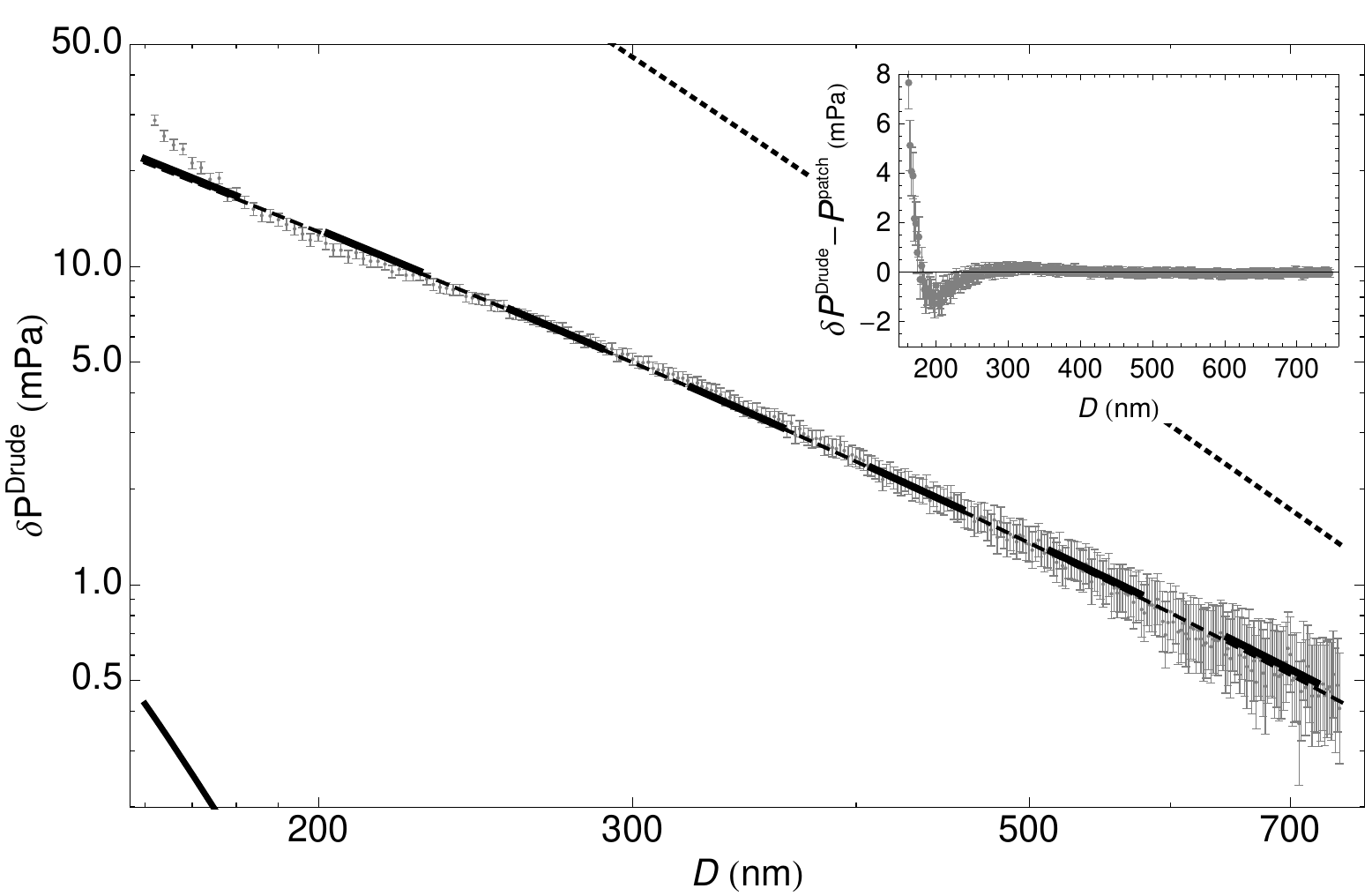}
\caption{ Comparison of the residual $\delta{P}^{\rm Drude}$ between
the experimental pressure in \cite{Decca07} and the Drude prediction
(points with error bars at $67\%$ confidence taken from Fig. 2 of \cite{Decca07}) with patch pressure $P_\mathrm{patch}$ for
four different patch models (more details in the main text)~: 1. The
solid curve is the result of the sharp-cutoff model (with
assumptions of subsection II-A)~; 2. The dotted curve corresponds to
the quasi-local patch correlation model assuming that the patch
sizes are given by the grain sizes and that the rms voltage is given
by the variance of the work function over different crystallographic
planes~;  3.  The long-dashed curve is the result of a best-fit on
the parameters ($\ell_\mathrm{max}$ and $V_\mathrm{rms}$) of the
quasi-local patch correlation model~; 4. The short-dashed curve
(underneath the long-dashed curve) is a fit of a phenomenological
model proposed in \cite{Carter11}. The inset shows the residual signal
resulting from subtracting the fit of the quasi-local model (long-dashed curve)
from $\delta P^{\rm Drude}$.}
\label{RicardoDataDrude}
\end{figure}

We collect in Fig.\ref{RicardoDataDrude} the information needed to
compare IUPUI experimental data with predictions from the Drude
model and modelings of the patch effect. We plot the residuals
$\delta{P}^{\rm Drude}$ defined as in (\ref{residuals}) as points
with error bars and the patch pressure $P_\mathrm{patch}$ for
different patch models as lines. The error bars represent the total
experimental error described in Fig. 2 of \cite{Decca07} at $67 \%$ confidence. 
The theoretical predictions for the Casimir pressure $P_{pp}^{\rm Drude}$ are calculated for the Drude model as described above and assumed to have no error. There are four different patch
models represented in Fig.\ref{RicardoDataDrude}~:
\begin{enumerate}

\item The solid curve is the estimation of the patch effect using
all the assumptions of subsection II-A. The patches are thus
described by the sharp-cutoff model (\ref{STPS}) with the parameters
$k_\mathrm{max} = 251 \, \mu\mathrm{m}^{-1}$, $k_\mathrm{min} =
20.9 \, \mu\mathrm{m}^{-1}$ and $V_\mathrm{rms} = 80.8$ mV
(these are the parameters used in \cite{Decca05}).

\item The dotted curve is the result of the quasi-local correlation
model (\ref{quasi-local-correlation}) with the patch size
distribution (\ref{SM}) described in subsection II-B. The
parameters, $\ell_\mathrm{patch}^\mathrm{min}=25$nm,
$\ell_\mathrm{patch}^\mathrm{max}=300$nm and
$V_\mathrm{rms}=80.8$ mV, correspond to the assumptions that the
patch sizes are given by the grain sizes and the rms voltage is
determined by the variance of the work function over the different
crystallographic planes (these are the same parameters used in item
1 above).

\item The long-dashed curve is obtained from a
least-squares minimization of the difference $ \delta{P}^{\rm Drude}
(D) - P^\mathrm{patch} (D)$, using the quasi-local patch correlation
model given by (\ref{quasi-local-correlation}) and (\ref{SM}). As
$\ell_\mathrm{patch}^\mathrm{min}$ is found to have a small influence, we fix it
to the smallest grain size $\ell_\mathrm{patch}^\mathrm{min}=25$ nm as discussed
above. The best fit on the two remaining parameters gives
$\ell_\mathrm{patch} ^\mathrm{max} \approx 2476$ nm and
$V_\mathrm{rms} \approx 9.2$ mV and results in qualitative agreement
between the residual and the fitted patch pressure. 
The associated first moments of the
patch size distribution are $\overline{\ell}= 1251$ nm and
$\overline{\ell^2}=(1437$ nm$)^2$. The reduced-$\chi^2$ for this fit,
calculated using the total error bars from Fig 2. of \cite{Decca07} at $67 \%$,
is $0.814$. It is important to note that the values of the fit parameters and quality of the fit are 
very sensitive to the sample's optical parameters, in particular to the plasma 
frequency used in the extrapolation of optical data to low frequencies \cite{sampledep}.
However,
one should avoid giving too much importance to any of these values of reduced-$\chi^2$ as a measure with statistical significance of experiment-theory agreement. Indeed, the influence of sample dependency of optical parameters, the use of a very crude description of roughness corrections to the Casimir  pressure, and, most importantly, the lack of precise information of the patch correlation
function in actual experimental samples, all imply that the fits obtained with the quasi-local model for patches have a qualitative nature; dedicated patch effects measurements are required to make metrological claims (see the Conclusions for further discussions).

\item The short-dashed curve (underneath the long-dashed curve)
is a fit of a phenomenological model proposed by Carter and Martin
\cite{Carter11}. The correlation function of this model, based on a
Monte-Carlo simulation of patch layouts, can be expressed in terms
of a shifted Gaussian and is specified by the rms voltage and the
average patch area $w^2$, related to our patch radius via $w
\approx \sqrt{\pi} \, \, \overline{\ell}/2$. Our best fit values are
$\overline{\ell} \approx 1229$ nm and $V_\mathrm{rms}= 8.6$ mV with
reduced-$\chi^2$ of 0.812.

\end{enumerate}

After this description of the information gathered on Fig.
\ref{RicardoDataDrude}, let us now comment on the significance of
the various results:

\begin{enumerate}
\item The solid curve reproduces and confirms the calculations
which were performed to quantify patch effects in
\cite{Decca05,Decca07}. With the assumptions described in subsection
II-A, the calculated patch pressure is indeed far too small to
explain the difference between experimental data and theoretical
predictions using the Drude model.

\item The dotted curve gives the result of the quasi-local
model of patch correlations (\ref{SM}) with parameters
determined as was done in \cite{Decca05,Decca07}, but here for a
different patch spectrum model. As a striking illustration of the
importance of this difference, the calculated patch pressure is now
larger than the difference between experimental data and
theoretical predictions using the Drude model. This illustrates the
 highly model dependent nature of the computed patch pressure. Thus, 
patches may be an {\it important systematic effect} for which their contribution
to the measured signal should ideally be assessed independently of any Casimir
force measurement. 

\item The long-dashed curve corresponds to a least squares fit of the
quasi-local correlation model to the residual $\delta P^{\rm
Drude}$. With the best-fit parameters
$\ell_\mathrm{patch}^\mathrm{max}$ and $V_\mathrm{rms}$, this model
qualitatively fits the difference between experimental data and
theoretical predictions using the Drude model. These parameters have
reasonable values: $\ell_\mathrm{patch}^\mathrm{max}$ is larger than
the maximum grain size on the samples, and $V_\mathrm{rms}$ smaller
than the rms voltage for a clean sample \cite{Decca05,Decca07}. This
suggests the presence of contaminants on the sample surfaces
\cite{Rossi92}.

\item The best-fit of the phenomenological model proposed in
\cite{Carter11} is essentially indistinguishable from that of the
quasi-local correlation model (long-dashed curve). The best-fit
values for $\overline{\ell}$ and $V_\mathrm{rms}$ are consistent with
the average patch size and rms
voltage obtained from the best-fit parameters of the quasi-local
model.
\end{enumerate}

At this point, we also want to comment on the validity requirement
(\ref{validity}), which allows one to calculate the patch effect in
the sphere-plane geometry within the PFA. This requirement ensures
that the effective area of interaction between the sphere and the
plane, of the order of $\pi RD$ for a sphere of radius $R$, contains
a large number of elementary patch areas, so that the sum over the
micro-realization of patches on a given plate is a good effective
description of the statistical ensemble-average given by the power
spectral density. With the numbers in \cite{Decca07}, that is a
radius of curvature of the sphere $R=151.3\mu$m and a shortest
distance $D_\mathrm{min}=160$nm, the interaction area is $\pi RD
\approx 76 (\mu\text{m})^2$. 
Meanwhile, the average patch area is
$(\pi/4)\overline{\ell^2} \approx 1.6 (\mu\text{m})^2$, 
there is a large number of elementary patch areas ($\approx 48$) within
the effective area of interaction, but it is possible that one could expect 
a small correction to the patch pressure at short distances when the ergodic 
hypothesis begins to break down.

\begin{figure}[t]
\centering
\includegraphics[width=3.3in]{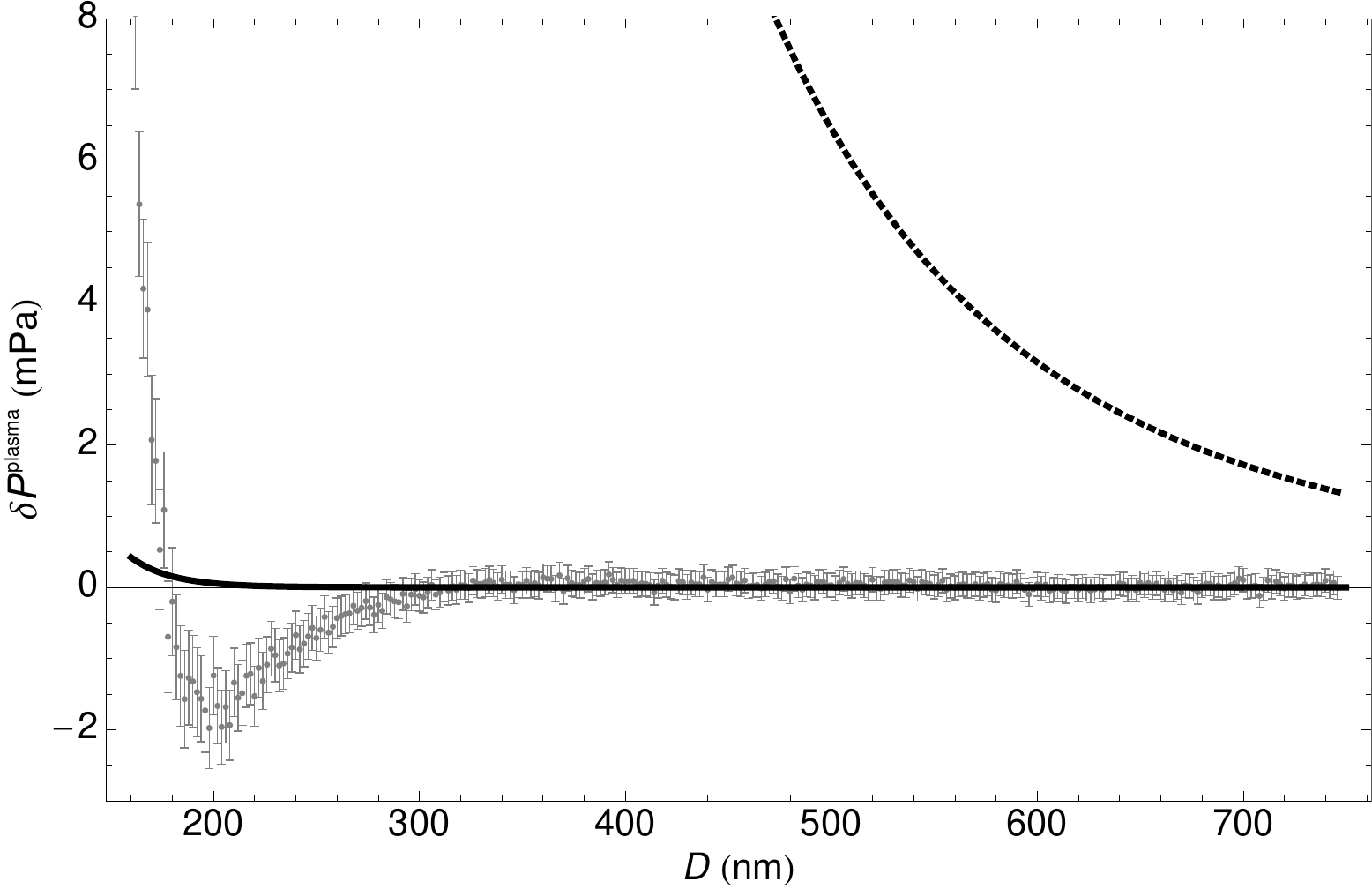}
\caption{ Comparison of the residual $\delta{P}^{\rm plasma}$ (points with error bars at $67 \%$ confidence taken from Fig. 2 of \cite{Decca07}) with
patch pressures given by the sharp-cutoff model and the quasi-local
model: 1. The solid curve is the result of the sharp-cutoff model
(with assumptions of subsection II-A).
We find, consistently with the analysis in \cite{Decca05,Decca07},
that the patch pressure from this model gives a negligible
contribution to the measured signal.
2. The dotted curve corresponds to the quasi-local patch correlation
model adopting the same parameters used in \cite{Decca05,Decca07}
for the patch size and the rms voltage.
In distinction to the sharp-cutoff model, we find that the
quasi-local model gives a large signal as compared to the residual
$\delta{P}^{\rm plasma}$. } \label{RicardoDataPlasma}
\end{figure}

For completeness we have also studied the residual $\delta P^{\rm
plasma}(D)$, as defined in Eq. (\ref{residuals}), with the exception
that we have compute the plane-plane Casimir pressure 
$P_{pp}^{\rm plasma}(D)$ using the ``plasma model",  instead of the Drude model. 
More precisely, we have computed the pressure using for the
permittivity $\varepsilon(i \xi)$ the
``generalized plasma model":

\begin{equation}
\label{generalizedplasma} 
\varepsilon^{\rm g. plasma}(i \xi)  = 1+ \frac{\Omega^2_P}{\xi^2} + \sum_{j=1}^6  \frac{f_j}{\omega_j^2 + \xi g_j
+ \xi^2} ,
\end{equation}
where the first two terms correspond to the permittivity for the
plasma model for conduction electrons (dissipation of conduction electrons is
set to zero ad hoc without physical justification), and the second sum of
terms accounts for the interband transitions of gold \cite{parameters}. 
To account for roughness corrections to the Casimir pressure we use
the same additive scheme employed above. Computing $\delta P^{\rm
plasma}(D)$ in this way, we have confirmed
the findings of \cite{Decca05,Decca07}, namely that a negligible
contribution of the patch effect leads to an agreement of data with
theoretical predictions using the plasma model. We note, however,
that the patch pressure calculated from the quasi-local model, with
sizes and voltages used in \cite{Decca05}, is much larger than the
difference between the measurements and the plasma prediction, as
shown in Fig. \ref{RicardoDataPlasma} \cite{footnote2}. We think
that this result constitutes a {\it serious warning} against the claims
according to which the plasma model would be confirmed with a high
confidence level by Casimir experiments performed with real metals
\cite{Klimchitskaya09}.

\subsection{Data analysis for the Yale experiment}

In addition to analyzing the IUPUI experiment we now apply the same
models to the recent experiment by the Yale group \cite{Sushkov11}.
A patch analysis was already carried out in \cite{Sushkov11} and it
led to a good agreement between experimental data and the Drude
model. This analysis only considered the asymptotic form $\propto
1/D$ of the plane-plane energy due to patches (\ref{largepatches}).
Here we extend the analysis by using the more general expression
(\ref{pressure}) for the patch pressure with the quasi-local patch
correlation function described in subsection II-B. Because we have
no information regarding grain or patch sizes in the Yale
experiment, we will focus our attention on best-fit estimations of
the parameters $\ell^{\rm max}_{\rm patch}$, $\ell^{\rm min}_{\rm
patch}$, and $V_{\rm rms}$ characterizing the quasi-local patch
correlation function (\ref{quasi-local-correlation},\ref{SM}).

\begin{figure}[t]
\begin{center}
\includegraphics[width=3in]{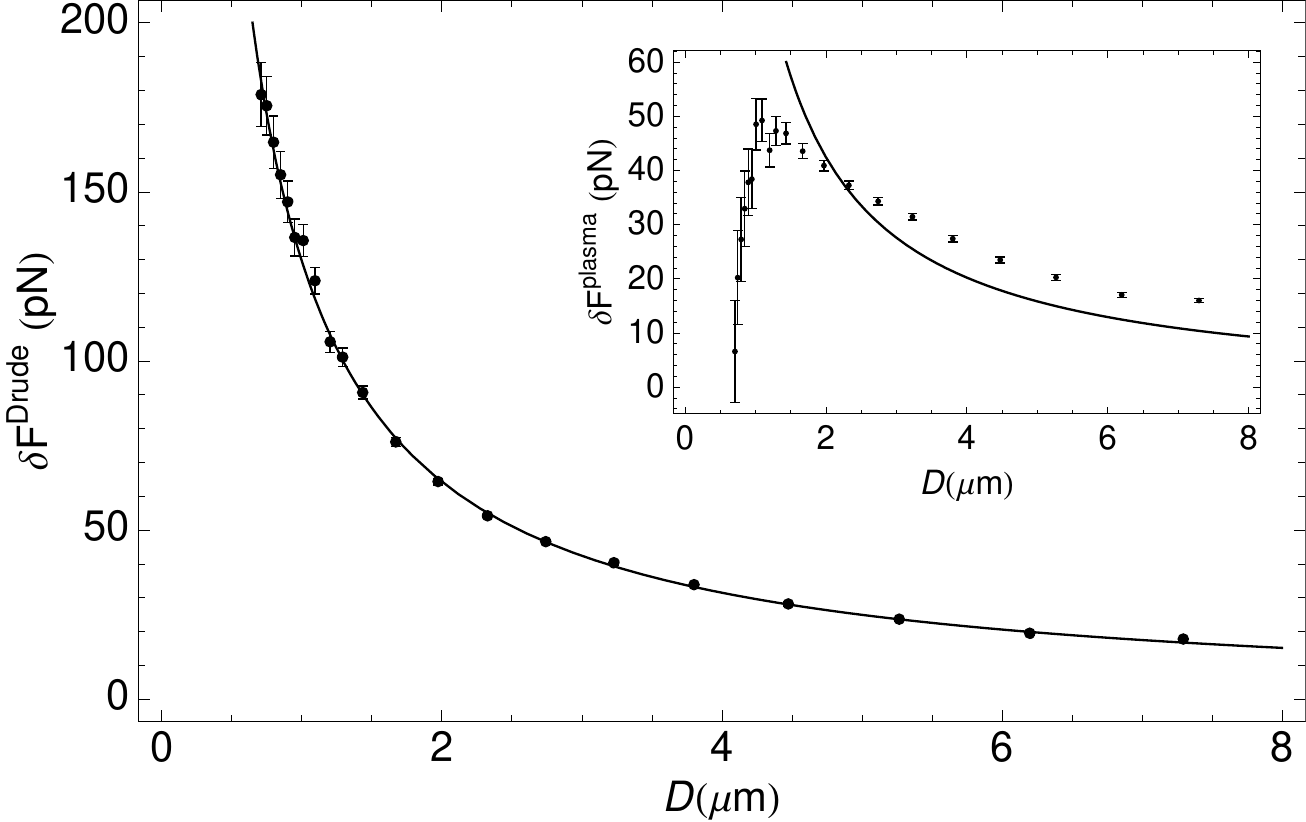}
\caption{ Comparison of the residual $\delta F^{\rm Drude}$ between
the data of \cite{Sushkov11} and the computed sphere-plane force
with the associated patch pressure $ F^\mathrm{patch}$. The dots
correspond to $\delta F^{\rm Drude}$, with the errors bars including
only the experimental error in the force determination. The solid
line is a best-fit of the patch force within the quasi-local model
of subsection II-B. The inset shows the corresponding residual
$\delta F^{\rm plasma}$ with the same convention employed in the
main figure. } \label{SteveFit}
\end{center}
\end{figure}

To analyze Yale experimental data we first compute the Casimir force 
using tabulated optical data extrapolated to low frequencies with the Drude
model using the plasma frequency $\Omega_P = 7.54$ eV and 
the dissipation rate $\gamma = 0.052$ eV employed in \cite{Sushkov11}.
We set the temperature to be $T=295$K.
The roughness correction to the Casimir  force is ignored as it gives a negligible
correction to the force at the distances considered in the Yale experiment.
Fig.\ref{SteveFit} shows the difference of the Yale experimental
force data and the Casimir force prediction using the Drude model,
$\delta F^{\rm Drude}$ (defined by analogy with (\ref{residuals})),
depicted by points with error bars (we assume no error for the
theory). The solid curve shows the resulting patch force for
parameters arising from a least-squares minimization of the quantity
$\delta F^{\rm Drude} - F^\mathrm{patch}$  using the quasi-local
patch correlation model (\ref{SM}) described in subsection II-B. The
best fit parameters are given by $\ell_\mathrm{patch} ^\mathrm{max}
\approx 614  \mu$m, $\ell_\mathrm{patch} ^\mathrm{min} \approx 566
\mu$m, (corresponding with $\overline{\ell}= 590 \mu$m) and
$V_\mathrm{rms} \approx 3.9$ mV. We should point out, however, that
the result of the best-fit is essentially insensitive to the
details of the patch power spectrum. Indeed, since the residual $\delta P^{\rm Drude}$ in the
Yale experiment has an approximate $1/D$ power law we can infer
using (\ref{largepatches}) that the typical patch size is much
larger than $D$ for the whole range of distances  explored in the
experiment (0.7 $\mu$m - 7$\mu$m). Performing a constrained fitting
by requiring that $\ell^{\rm max}_{\rm patch}$ be less than some
predetermined value (e.g. 500 $\mu$m), yet still satisfying the
constraint $\bar{\ell} \gg D$, we were able to verify that  a good
fit can still be achieved over a large range of patch sizes. In
summary, we point out the result of our fitting using the more
detailed quasi-local patch model confirms the patch treatment in
\cite{Sushkov11}.

Finally, we also report for the sake of completeness some
supplementary test we performed for comparing the data in
\cite{Sushkov11} with the predictions of the plasma model  described by Eq.(\ref{generalizedplasma})
(see the inset of Fig. \ref{SteveFit}). The good agreement obtained for the
Drude model is dramatically degraded. Therefore, we confirm the
result obtained in \cite{Sushkov11} that patches cannot explain the
difference between the experimental data and the plasma model in
Yale data.


\section{Concluding Remarks}

In this paper, we have analyzed the patch contribution to Casimir
experiments with a model featuring quasi-local voltage correlations.
Our model is derived from well-motivated physical principles and shares
key features with experimentally verified patch models used to describe
ion trap heating and cantilever damping \cite{Dubessy09}.
Thus, for the description of the surfaces used in the experiments discussed in this paper, we
believe that this model is more appropriate than the sharp-cutoff
model which has been used to the same aim in previous publications
\cite{Decca05,Decca07}. 

Due to the large difference in the patch
power spectrum, in particular for small wavevectors, the quasi-local
model gives a larger contribution than the sharp-cutoff model. As a
striking consequence, when the patch sizes are deduced from the
grain sizes (as was done in \cite{Decca05,Decca07}), the quasi-local
model produces a patch pressure larger than the difference between
the experimental data and the Drude (and plasma) Casimir prediction,
whereas the sharp-cutoff model produces a negligible patch pressure.
Therefore, it is important to emphasize that because of the combination
of: a)  the highly model dependent nature of the computed patch pressure 
and b) the potentially large patch contribution to the 
measured signal, patches may lead to nonnegligible systematic effects. 
This necessitates an independent  measurement of patch effects in 
order to meet metrological standards for Casimir force measurements. 

We have also used the new quasi-local patch model to fit the difference between experimental data of the 
IUPUI experiment \cite{Decca05,Decca07} and the theoretical prediction for the Casimir pressure. The latter
was computed taking into a) tabulated optical data extrapolated to low frequencies by means of the 
Drude model, and b) roughness effects modeled by a simple additive technique.
We have
found  best-fit parameters for the
average patch size and for the rms voltage that are consistent with
a contamination of the metallic surfaces, which is expected to
enlarge the patch sizes (with respect to grain sizes) and smear the
patch voltage (with respect to those of a surface of bare
crystallites) \cite{Rossi92}. 
Indeed, surface contamination is expected, and we believe that preferential adsorption \cite{Rossi92} and saturation of contaminants 
may be compatible with the observation of reproducible results in experiments repeated several times with different samples \cite{DeccaPC}.

Taken together, our results constitute a {\it strong warning} against
the previously published claims of an agreement of Casimir
experiments with the plasma model, and an elimination of the Drude
model \cite{Klimchitskaya09}. However,  we want to emphasize that
they do not constitute yet a proof of agreement of experimental data
with the new model. The parameters of the patch model have been
fitted and it is still possible that the qualitative agreement thus
obtained is a fortunate output of the fitting procedure rather than
an explanation of the experimental data. 

In this paper we have focused our attention
on only the IUPUI and Yale experiments, but of course the analysis can be
repeated for other Casimir measurements between metallic plates as
well \cite{Chan01,Bressi02,Chen04,Lisanti05,Svetovoy08,Onofrio08,Jourdan09,deMan09,Masuda09,Antonini09}.

A better characterization of the surfaces used in the experiments is
now key to reaching firmer conclusions. The patch distributions can
be measured with appropriate technologies such as Kelvin probe force
microscopy which can achieve the necessary size and voltage
resolutions \cite{Liscio08,Liscio11}. In addition, the study of cold
atoms and cold ions trapped in the vicinity of metallic surfaces
\cite{Epstein07} or the role of patch effects in other precision
measurements \cite{Adelberger09,Everitt11,Reasenberg11} are other
ways for accessing information of interest for our problem. Let us
repeat at this point that our new quasi-local model is similar to
recent proposals for patch physics used to achieve a better
understanding of atomic and ionic traps \cite{Dubessy09,Carter11}.

The challenges of forthcoming studies may be stated as follows.
First, it is important to confirm the hypothesis that the patch
voltages show quasi-local correlations, and to better specify the
power spectrum which quantitatively describes these correlations.
Second, it would also be interesting to study how the patch power
spectrum depends on contamination, in particular, fabrication,
treatment, history of the samples, and on temperature. Finally, an
independent determination of the patch power spectrum could lead
either to a confirmation of the best-fit analysis presented in this
paper or to new questions. This study is important not only for the
test of the Casimir effect, a central prediction of quantum field
theory, but also for the searches of the hypothetical new
short-range forces predicted by unification models
\cite{Fischbach98,Adelberger03,Onofrio06,Antoniadis11}.

\acknowledgments

We are grateful to Ricardo Decca, Steve Lamoreaux, Alex Sushkov, and
Woo-Joong Kim for having kindly provided experimental data and
information needed to analyze them, and for many insightful
discussions. We also acknowledge discussions with Astrid Lambrecht,
Antoine Canaguier-Durand, Giovanni Carugno, Jo\"el Chevrier, Thomas Coudreau, Thomas
Ebbesen, Cyriaque Genet, Romain Gu\'erout, Harald Haakh, Carsten Henkel, Galina Klimchitskaya, Johann
Lussange, Sven de Man, Umar Mohideen, Vladimir Mostepanenko, Roberto Onofrio, Giuseppe Ruoso, Paolo Samori, Signe
Seidelin, and Clive Speake.

This work was supported by the US Department of Energy through contract
DE-AC52-06NA25396 and was partially funded by LANL LDRD program and by
DARPA/MTO's Casimir Effect Enhancement program under DOE/NNSA
Contract DE-AC52-06NA25396. P. A. M. N. thanks CNPq and FAPERJ-CNE
for partial financial support. The authors are thankful for the ESF
Research Networking Programme CASIMIR (www.casimirnetwork. com) for
providing excellent opportunities for discussions on the Casimir
effect and related topics.

\newcommand{\Review}[1]{{\em #1}}
\newcommand{\Volume}[1]{\textbf{#1}}
\newcommand{\Book}[1]{\textit{#1}}
\newcommand{\Eprint}[1]{\textsf{#1}}
\def\etal{\textit{et al}}


\begin{references}
\bibitem{footnote}
As a minor remark, we should point out that the definitions of
$V_\mathrm{rms}$ used here and in \cite{Sushkov11} differ by a
factor $\sqrt{2}$.
\bibitem{parameters} The values of the interband parameters used in Eq.(\ref{generalizedplasma})
are:
$\omega_1 = 3.05$, $g_1 = 0.75 $, $f_1= 7.091$; 
$\omega_2 = 4.15$, $g_2 = 1.85 $, $f_2= 41.46$; 
$\omega_3 = 5.4$, $g_3 = 1.0$, $f_3= 2.7$; 
$\omega_4 = 8.5$, $g_4 = 7.0$, $f_4= 154.7$; 
$\omega_5 = 13.5$, $g_5 = 6.0$, $f_5= 44.55$; and
$\omega_6 = 21.5$, $g_6 = 9.0$, $f_6= 309.6$.
Units of $\omega_j$ and $g_j$ are eV, units of $f_j$ are $({\rm eV})^2$.
\bibitem{footnote2}
The quasi-local model can also be fitted to the residual $\delta
P^{\rm plasma}$, but, not surprisingly, the fit is insensitive to
patch sizes as the residual signal is compatible with 0 over a large
range, implying $V_{\rm rms} \approx 0$.
\end{references}
 \end{document}